\documentclass[12pt]{iopart}
\usepackage[top=35truemm,bottom=20truemm,left=35truemm,right=35truemm]{geometry}

\usepackage{iopams} 
\usepackage[dvipdfmx]{graphicx}% Include figure files
\expandafter\let\csname equation*\endcsname\relax
\expandafter\let\csname endequation*\endcsname\relax
\usepackage{amsmath}
\usepackage{textcomp}
\usepackage{color}
\usepackage{ulem}
\usepackage{cite}
\usepackage{textcomp}
\usepackage{multicol} 

\begin{document}

\title[]{Sensitivity Improvement by Sample Vibration Excitation in Resistivity Measurement for Non-Magnetic Material Using MFM}
%アナリティかるに変える、構図変える

\author{Kazuma Okamoto$^1$, Takumi Imura$^1$, Naruto Nakamura$^1$, Satoshi Abo$^1$, 
Katsuhisa Murakami$^2$, Masayoshi Nagao$^2$, Fujio Wakaya$^1$}

\address{$^1$ School / Graduate School of Engineering Science, The University of Osaka, 1-3 Machikaneyama, Toyonaka, Osaka 560-8531, Japan}
\address{$^2$ National Institute of Advanced Industrial Science and Technology, 1-1-1 Umezono, Tukuba, Ibaraki 305-8568, Japan}
\ead{wakaya.f.es@osaka-u.ac.jp}
\vspace{10pt}
\begin{indented}
\item[]9 December 2025
\end{indented}
\begin{abstract}
A novel approach for measuring the electrical resistivity of non-magnetic materials 
using magnetic force microscopy (MFM) is discussed. 
In this method, MFM detects magnetic fields generated by eddy currents induced by 
the oscillation of a magnetized probe tip. 
To enhance measurement sensitivity, it is essential to increase the magnitude of these eddy currents.
It is discussed 
that introducing controlled sample vibration amplifies 
eddy current generation by increasing the relative velocity 
between the probe tip and the sample surface.
Theoretical analysis predicts increase of the phase shift by sample vibration, 
and experimental validation using a modified MFM system 
confirms the improvement in sensitivity.
The calculated and experimental results exhibit relatively good agreement, 
establishing that sample vibration excitation is an effective strategy for high-sensitivity resistivity measurements.

\end{abstract}

\ioptwocol

\section{Introduction}
The impurity density or resistivity of semiconductors is a 
critical parameter for semiconductor devices\cite{dk}. 
Various doping techniques---such as thermal diffusion, 
ion implantation, 
plasma doping, 
and epitaxial growth---are employed to introduce 
impurities\cite{cz,cx,da}. 
Post-doping annealing is commonly used to activate these impurities, but it can also cause atomic displacement, necessitating precise characterization methods\cite{bp,db,dc}. 

Scanning spreading resistance microscopy (SSRM) offers high spatial resolution for impurity profiling by measuring current under an applied voltage\cite{cx}. 
However, its reproducibility is limited due to surface scratching\cite{cv}. 
Magnetic force microscopy (MFM) is generally used for imaging 
magnetic domain structures, 
where a magnetized tip oscillating near the sample surface is used\cite{an,dd,br}. 
In this study, an impurity density measurement method using MFM is discussed. 
The reproducibility of such MFM measurements is better than that of SSRM because the tip and sample surface are neither touched nor biased\cite{df}. 

It has been reported that 
MFM can detect signals from non-magnetic materials by sensing magnetic fields generated by eddy currents induced by tip oscillation\cite{ci,n,dt,du}. 
The phase shift in tip oscillation, proportional to the derivative of the magnetic field, enables resistivity estimation\cite{n}. 
Given that semiconductor resistivity is significantly higher than that of metals \cite{cs,af,ea,eb,ec}, improving MFM sensitivity is essential\cite{n}.  
It is hypothesized that 
sample vibration excitation, by increasing tip-sample relative velocity, enhances eddy current generation and thus measurement sensitivity.

This work aims to theoretically and experimentally validate the effectiveness of sample vibration excitation in improving resistivity measurement sensitivity using MFM.

\section{Theory}
\begin{figure}[t]
\centering
\includegraphics[width=80mm]{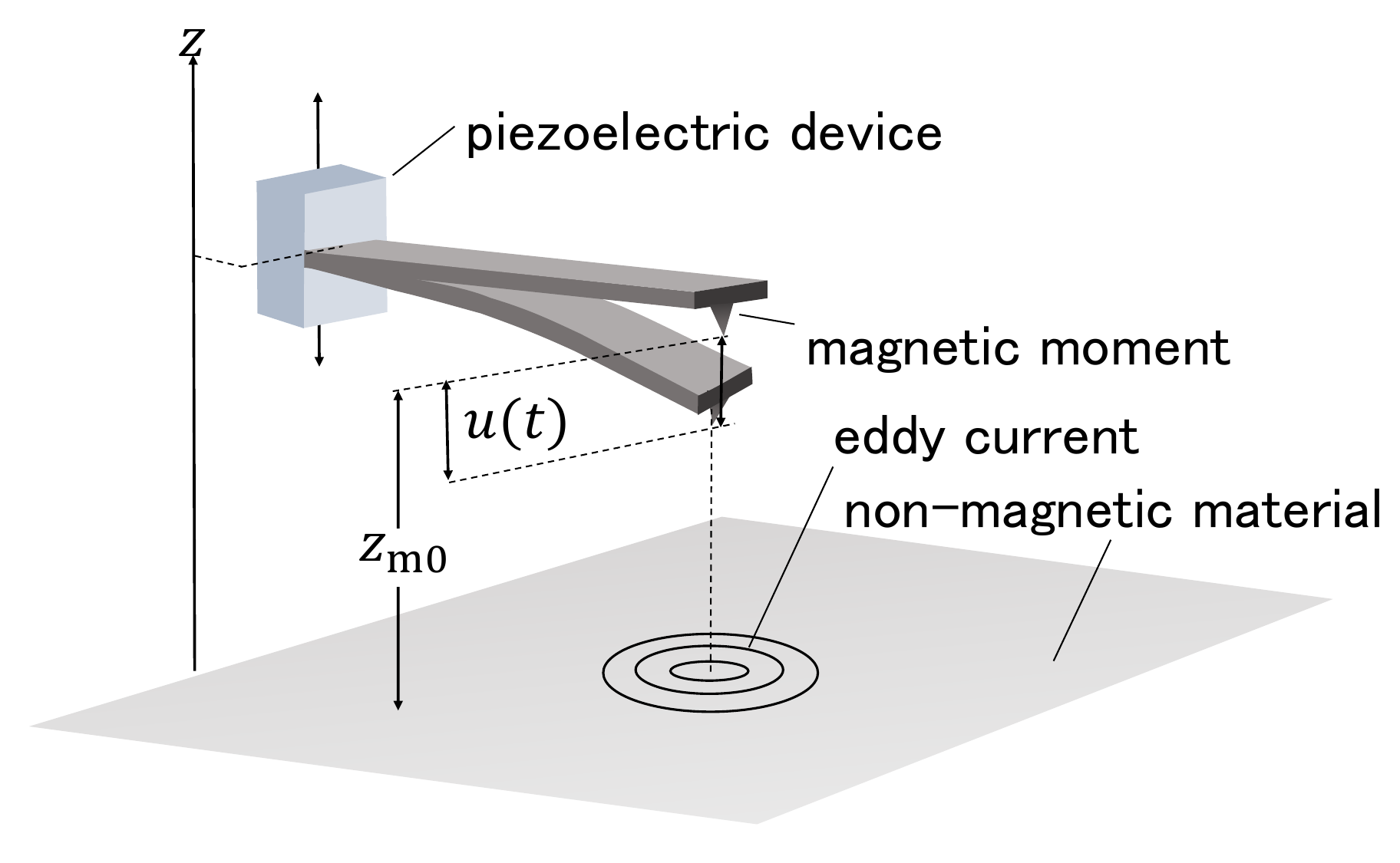}
\caption{Schematic of MFM system measuring resistivity of non-magnetic material. $u(t)$ is the displacement of tip and $z_{\rm m0}$ is the center height of tip oscillation.}\label{fig:model}
\end{figure}

\begin{figure}[t]
\centering
\includegraphics[width=90mm]{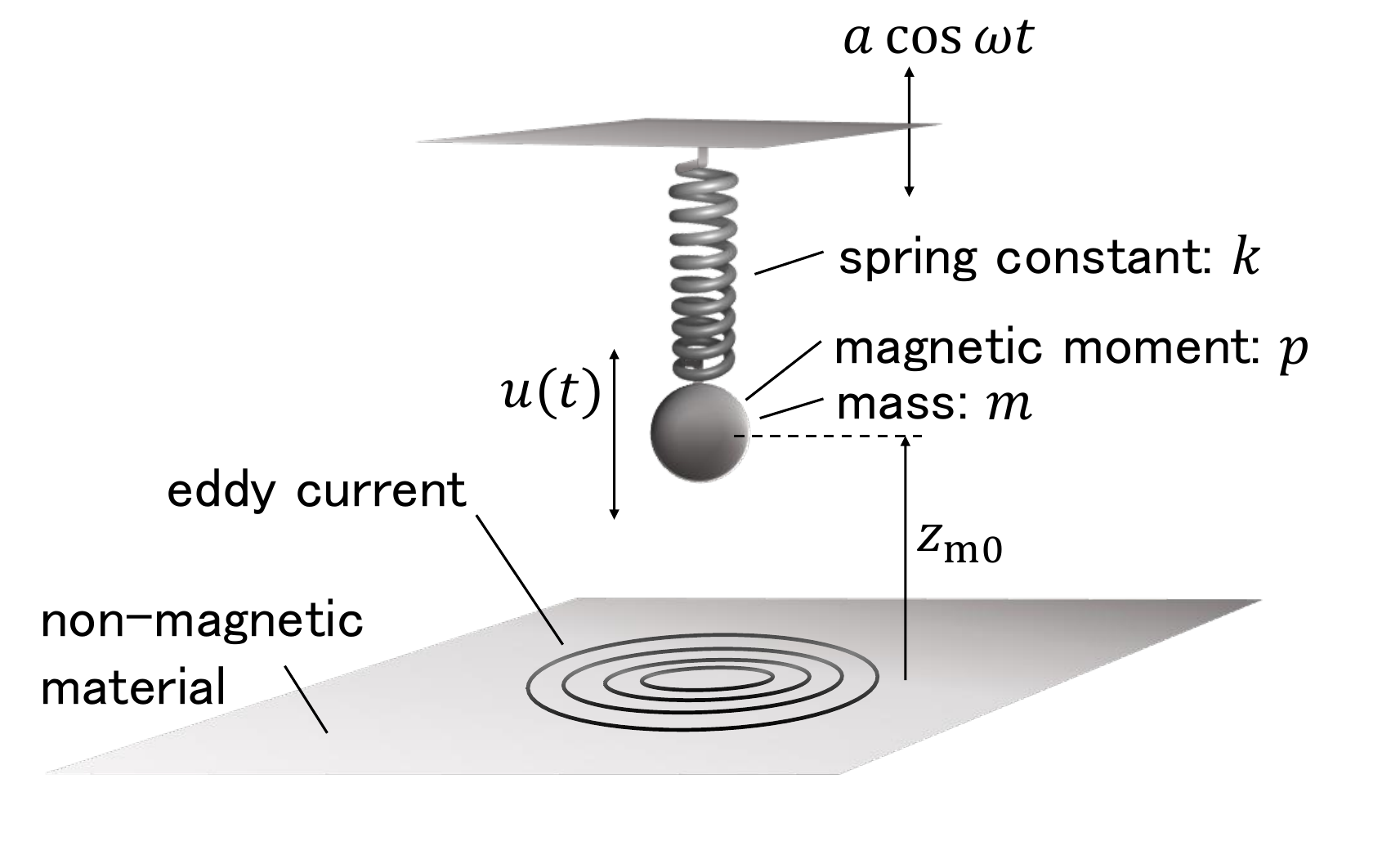}
\caption{Schematic equivalent to Fig. \ref{fig:model} for analyzing $u(t)$. $a$ is the amplitude of piezoelectric device, $\omega$ is the excitation frequency of tip, $k$ is the spring constant of cantilever, $p$ is the magnetic moment of tip and $m$ is the effective mass of tip.}\label{fig:modelspring}
\end{figure}

Figure \ref{fig:model} illustrates the MFM system for resistivity measurement of non-magnetic materials, 
where $u(t)$ and $z_{\rm m0}$ denote the displacement of the tip and the average height of the oscillating tip from the sample surface, respectively. 
Figure \ref{fig:modelspring} shows the equivalent mechanical model,  
where $a$, $\omega$, $k$, $p$ and $m$ denote the amplitude of the piezoelectric device, excitation frequency of the tip, spring constant of the cantilever, effective magnetic moment of the tip at a distance above the tip end, and effective mass of the tip, respectively\cite{ai,ct}. 
The tip is replaced with a spring and point mass for analyzing $u(t)$. 
%In this section, the differential equation and numerical analysis method are described.

%\subsection{{\rm Differential equation of the tip}}
The governing differential equation for tip motion under sample vibration excitation is expressed as
\begin{align}
&m\ddot{u}(t)+\frac{m\omega_0}{Q}\dot{u}(t)+ku(t)= \nonumber\\
&ak\cos\omega t-\frac{3p^2}{64\pi\rho}\frac{\dot{u}(t)+\omega\tilde{A}\sin(\omega t+\phi_{\rm s})}{\{z_{\rm m0}+u(t)-\tilde{A}\cos(\omega t+\phi_{\rm s})\}^4}+F_I,\label{eq:tip}
\end{align}
which is modified from the equation in Ref.\cite{n}.
Here, $Q$, $\rho$, $\tilde{A}$, and $\phi_{\rm s}$ denote the quality factor, sheet resistivity of the sample, amplitude and phase of the sample vibration, respectively. 
The second term on the left-hand side of Eq.~(\ref{eq:tip}) signifies dissipation. 
The dissipation coefficient is given by $m\omega_0/Q$\cite{ct} as is well known, where $\omega_0 \equiv \sqrt{k/m}$. % is the resonance angular frequency. 
The second term on the right-hand side of Eq.~(\ref{eq:tip}) is the magnetic force generated by the eddy current. The third term on the right-hand side of Eq.~(\ref{eq:tip}) is the interaction force that depends only on the tip-sample distance. The interaction force is modeled as: 
\begin{align}
&F_I(t)=\nonumber\\
&
\begin{cases}
%F_{vdW}(t)= %\nonumber\\
%&
-\frac{A_{\rm H}R}{6(z_{\rm m0}+u(t)-d)^2}\hspace{20pt}(a_0<z_{\rm m0}+u(t)-d)\\
%&
%F_{DMT}(t)=
-\frac{A_{\rm H}R}{6a_0}+\frac{4}{3}E^*\sqrt{R} %\nonumber\\
%&
(a_0-(z_{\rm m0}+u(t)-d))^\frac{3}{2}\\
\hspace{95pt}(z_{\rm m0}+u(t)-d\leq a_0)
\end{cases}
\end{align}
where $A_{\rm H}$, $R$, $d$, $a_0$ and $E^*$ represent the Hamaker constant, 
tip radius, the distance between the tip end and magnetic moment position, 
intermolecular distance 
and effective elastic modulus, respectively\cite{dz,dy}. 

It is difficult to derive analytical solutions of Eq. (\ref{eq:tip}) because it contains nonlinear terms. To solve Eq. (\ref{eq:tip}), therefore, the fourth-order Runge-Kutta method is used \cite{dw,dx} with initial values of $u(t)$ and $\dot{u}(t)$ of zero. 
The steady-state solution of Eq. (\ref{eq:tip}) was obtained after some time. 

\begin{figure}[t]
\centering
\includegraphics[width=80mm]{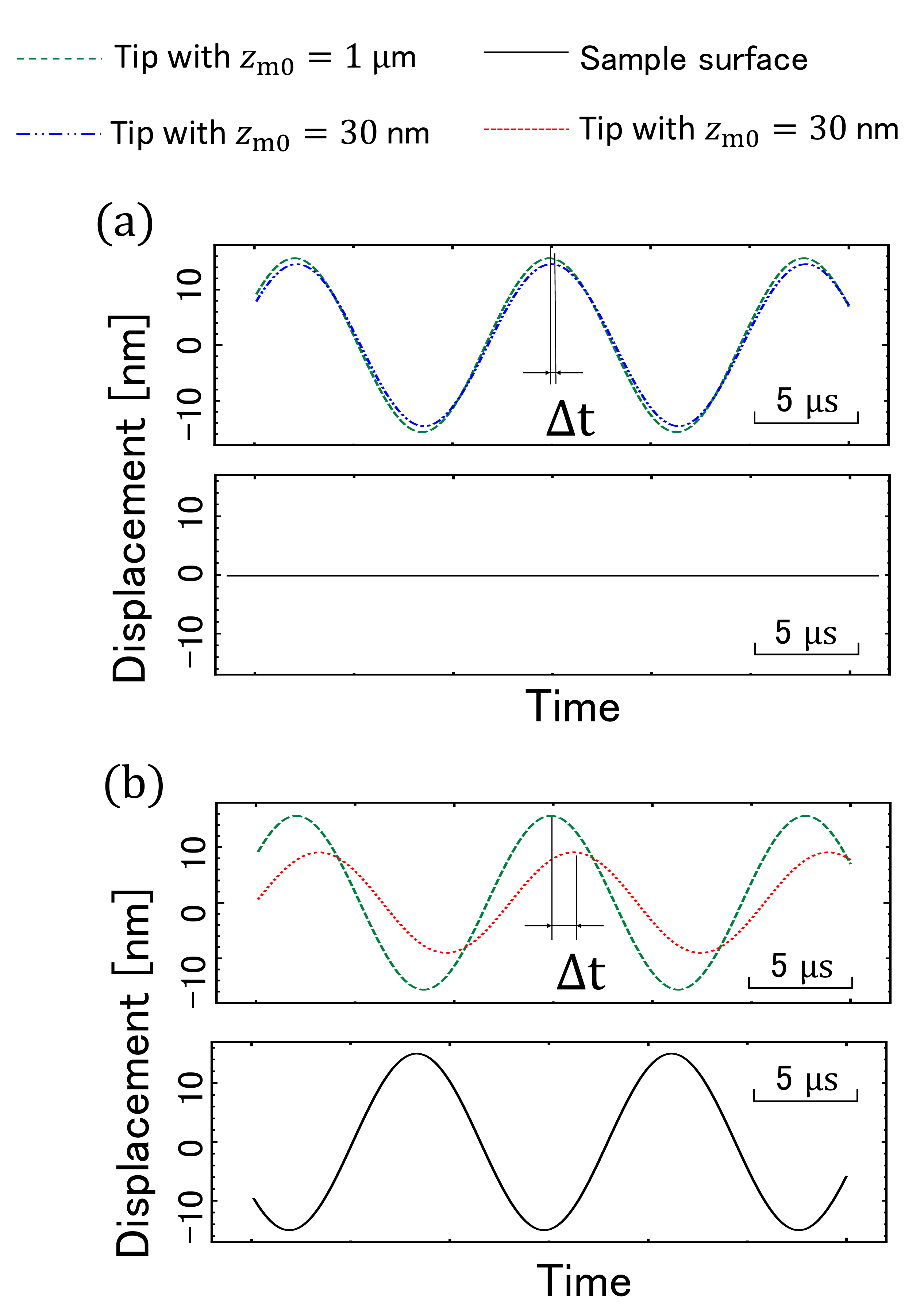}
\caption{The numerical solution $u(t)$ (a) without and (b) with sample vibration excitation where $a=0.1\;$nm, $p=1.0\times10^{-18}\;{\rm Wb\cdot m}$, $f_0=78\;$kHz, $f=0.998\times f_0$\;kHz, $k=3.0\;$N/m, $Q=200$, $\rho=10^3\;\Omega$/sq, $\tilde{A}=15.0\;$nm and $\phi_{\rm s}=3\pi/4$, $A_{\rm H}=0.0$\;J.}\label{fig:u}
\end{figure}
Figure.~\ref{fig:u} shows a typical example of such a steady-state solution.
The green and blue lines in Fig.~\ref{fig:u}(a) represent 
$u(t)$ with $z_\mathrm{m0}= 1$\;\textmu m and 30\;nm, respectively. 
The red line shows $u(t)$ 
with sample vibration excitation and $z_\mathrm{m0}=30$\;nm. The black lines represent the positions of the sample. The green line in Fig. \ref{fig:u}(b) is the same as that in (a) because 
the tip is sufficiently far from the sample as 1\;\textmu m and is not affected by eddy currents. 
The $\Delta t$ in Fig.~\ref{fig:u} is 
the difference in time when the tip displacement reaches its peak. 
The phase difference $\Delta\phi$ can be derived as $\Delta\phi=\omega\Delta t$. 

\section{Experimental procedure}
%The conventional MFM measures a oscillation-phase difference at each sample point with 
%scanning the proble horizontally. 
%When comparing the phase difference without a sample vibration and 
%that with a sample vibration, moving the probe horizontally is inconvenient.
%This is because the closest distance between the tip and the sample varies according to whether the sample is vibrated. The sensitivity of this measurement is inversely proportional to the fourth power of the distance between the tip and the sample. In this experiment, therefore, MFM measures the phase difference by moving the probe vertically. 

\subsection{{\rm Exciting sample vibration}}
\begin{figure*}[t]
\centering
\includegraphics[width=110mm]{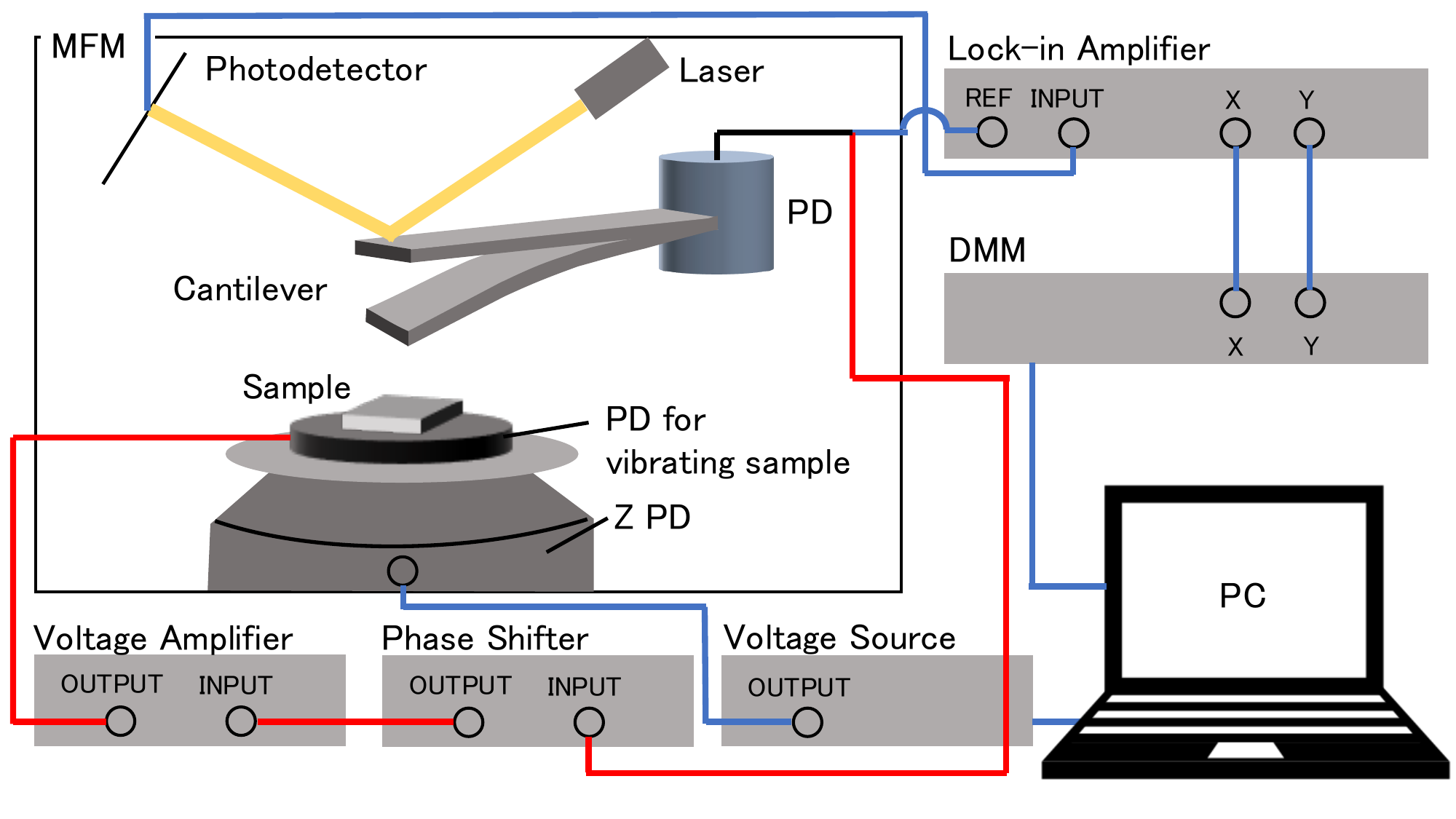}
\caption{Schematic drawing of measuring system for non-magnetic material.}\label{fig:equipment}
\end{figure*}

The MFM equipment was modified to vibrate the sample. 
Figure \ref{fig:equipment} illustrates the measuring system for non-magnetic materials used in this experiment.
The most significant change was the placement of a piezoelectric device 
beneath the sample. The sample was vibrated by the piezoelectric device. 
To match exactly the excitation frequency of the sample vibration to that of the cantilever vibration, the voltage of the piezoelectric device for the cantilever was divided and provided to the piezoelectric device for the sample. 
Before applying the voltage to the piezoelectric device for the sample, its phase and amplitude were adjusted using the phase shifter and voltage amplifier as shown in Fig.~\ref{fig:equipment} to achieve good sensitivity. 
%In the phase shifter, the sinusoidal waveform devided from the voltage applied at the piezoelectric device for the cantilever is initially converted to a square wave using a comparator. Then, the voltage phase is shifted by a digital phase shifter, which can shift the phase by up to 360$^\mathrm{o}$. Using a band-pass filter, 
%the square wave is converted to a sinusoidal wave. Finally, the voltage amplitude was amplified using a voltage amplifier. \\

\subsection{{\rm Phase difference measurement}}
To measure phase difference with varying $z_\mathrm{m0}$, a voltage controlled by the PC was applied to the Z piezoelectric device. 
The phase of the tip oscillation was varied by the forces from the sample. 
The photodetector transferred the tip oscillation to the voltage oscillation. 
The lock-in amplifier detects the phase difference between the oscillations of the tip and the voltage applied to the piezoelectric device for the cantilever. 
This phase difference was recorded in the PC. 
%This operation was performed with varying the tip height.

\section{Results and discussion}
\subsection{{\rm Phase difference}}
\begin{figure}[t]
\centering
\includegraphics[width=70mm]{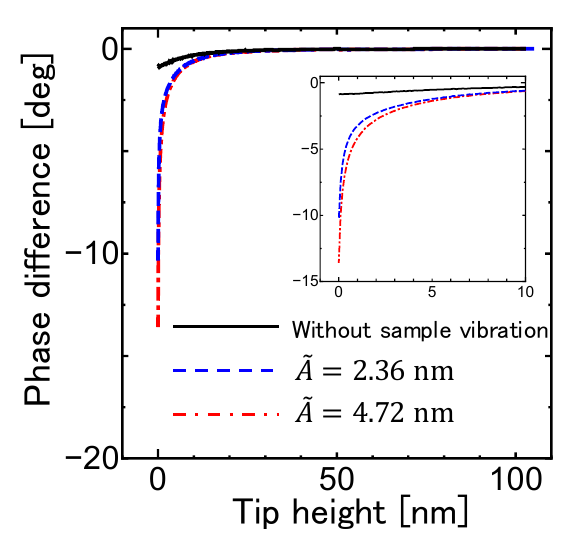}
\caption{Numerical solutions of phase difference where $a=0.2\;$nm, $f_0=63.129$\;kHz, $f=0.998\times f_0$\;kHz, $k=3.0$\;N/m, $Q=200$, $p=1.0\times10^{-18}$\;Wb$\cdot$ m, $\rho=7.7\;\Omega$/sq, $\phi=3\pi/4$, $A_{\rm H}=2.96\times10^{-19}$\;J, $R=40\;$nm, $a_0=4\;$${\rm \AA}$, $E^*=10.2\;$GPa and $d=15\;$nm\cite{dy,ed,ee,ef,eg,eh,en,eo}. The inset is an enlarged portion of the graph.}\label{fig:takasa}
\end{figure}

Figure \ref{fig:takasa} shows the phase difference $\Delta\phi$ derived by the numerical analysis as a function of $z_{\rm m0}$ where the black line is a phase difference without a sample vibration, the blue and red lines are phase difference with a sample vibration. 
As the tip approaches the sample, the phase difference increases negatively. Additionally, by increasing the amplitude of the sample oscillation, the phase difference increased negatively. This is caused by an increase in the relative speed between the tip and the sample. 
%An increase in the phase difference means an easy measurement. 
The increase in phase difference by sample vibration leads to 
the resistivity measurement of high-resistance samples, such as semiconductors. 

\begin{figure}[t]
\centering
\includegraphics[width=70mm]{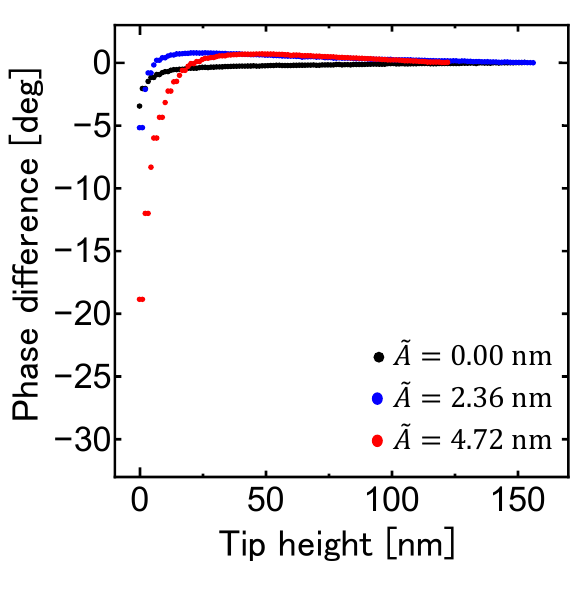}
\caption{Experimentally observed phase difference where $f_0=67.562\;$kHz, $f=67.3455\;$kHz, $k\simeq3.0\;$N/m, $Q=110.127$, $\rho = 7.7\;\Omega$/sq and $\phi_{\rm ts} = \pi$. $\phi_{\rm ts}$ is the phase difference between the tip and sample.}\label{fig:tildeA}
\end{figure}

To validate the calculation results, the experiments using 15-nm-thick gold thin film on a ${\rm SiO_2}$ substrate was performed. The sheet resistivity of the thin film was 7.7 $\Omega/$sq that was measured using a four-point probe\cite{ec,dv}.
The increase of negative phase difference was successfully observed by 
the experiments as shown in Fig.~\ref{fig:tildeA}, where 
the amplitudes of the sample vibration were fixed at specific values. 
The phase difference increased negatively with an increase in the sample vibration amplitude in the experiments as well as the calculation results shown 
in Fig.~\ref{fig:takasa}.
%As the amplitude of a sample is 2.36\;nm, 
%the phase difference is larger than ten times 
%the phase difference without a sample vibration. 
%As the amplitude of a sample is 4.72\;nm, 
%the phase difference is larger than twenty times 
%the phase difference without a sample vibration. 
%It is meaningless to compare the absolute values of the numerical analysis results %and the experimental results because this measurement has many parameters, and it is difficult to match the results of the numerical analysis and the experiment. 

\subsection{{\rm Origin of observed phase difference}}

It is necessary to consider effects other than eddy currents on the phase difference, such as the propagation of the sample vibration to the tip through the MFM body, $z_\mathrm{m0}$-dependence of the air resistance, etc. 
Control experiments were, therefore, performed.  

\begin{figure}[t]
\centering
\includegraphics[width=70mm]{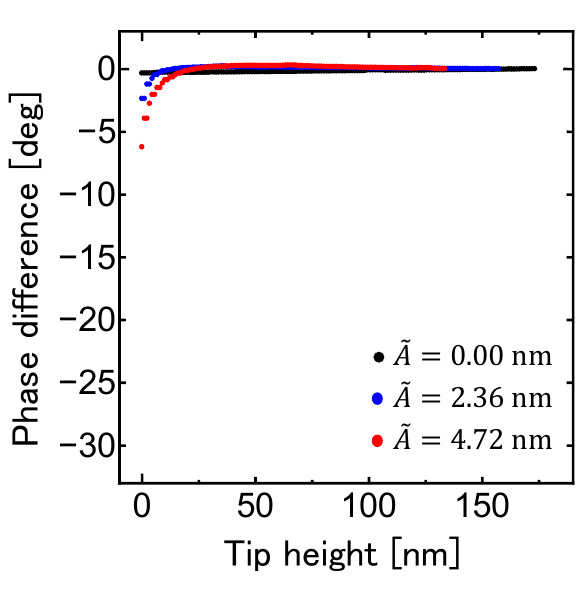}
\caption{Experimental result of a phase difference which is measured by a non-magnetized tip where $f_0=67.545\;$kHz, $f=67.3797\;$kHz, $k\simeq3.0\;$N/m, $Q=144.246$, $\rho=7.7\;\Omega$/sq and $\phi_{\rm ts}=\pi$.}\label{fig:nop}
\end{figure}

Figure~\ref{fig:nop} shows the phase difference measured using gold as a sample and a tip that was not intentionally magnetized. By using the non-magnetized tip, the effect of the eddy current can be eliminated. 
%%%%%%%%%%%%%%%%%%%%%%%%%%%%%%%%%%%%%%%%%%%%%%%%%%%%%%%%%%%%%%%%%%%%%%%%%
As shown in Fig.~\ref{fig:nop}, the phase difference measured using the non-magnetized tip is much smaller than that measured using the magnetized tip. 
The cause of the measured slight phase difference in Fig.~\ref{fig:nop} is 
that the tip is unintentionally magnetized slightly 
or effects other than eddy currents.

\begin{figure}[t]
\centering
\includegraphics[width=70mm]{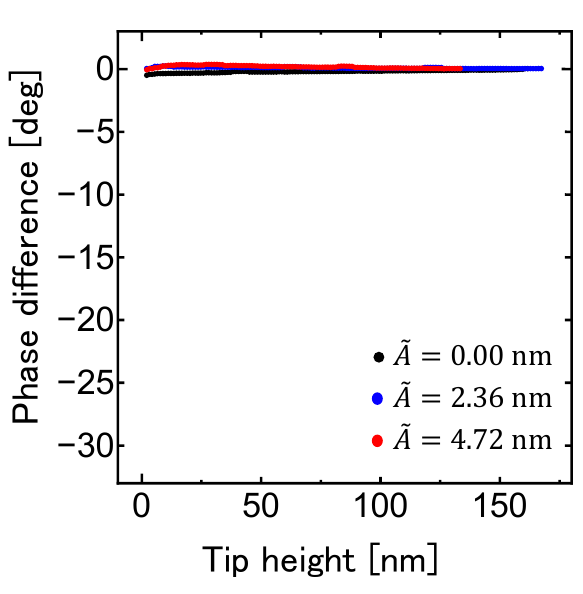}
\caption{Experimentally observed phase difference measured using ${\rm SiO_2}$ where $f_0=67.549\;$kHz, $f=67.3758\;$kHz, $k\simeq3.0\;$N/m, $Q=140.040$, $\rho=7.7\;\Omega$/sq and $\phi_{\rm ts}=\pi$.}\label{fig:SiO2}
\end{figure}

Another control experiment using SiO$_2$ was, therefore, performed. 
Figure~\ref{fig:SiO2} shows the phase difference measured using ${\rm SiO_2}$ as the sample and the magnetized tip. Eddy currents cannot be generated in ${\rm SiO_2}$. The effects other than eddy currents are, therefore, measured in this experiment.
As shown in Fig.~\ref{fig:SiO2}, the observed phase difference is almost zero, which means that the effects other than eddy currents do not affect the phase difference. 

These results show that the increase in eddy current accounts for 
%the majority of 
the major cause of the observed increase in the phase difference.

\section{Conclusions}
To improve the sensitivity in measuring resistivity by using MFM, 
a new technique using sample vibration excitation was proposed. 
The numerical solution showed that 
the sensitivity should be improved by vibrating the sample. 
To demonstrate the sensitivity improvement, the MFM was equipped with a piezoelectric device for sample vibration, a phase shifter, and a voltage amplifier. 
The phase difference with the sample vibration was actually detected 
using the modified MFM. 
The experimental results show that the sensitivity of the measurement was improved by sample vibration. Comparing the measurements with and without sample vibration, the sensitivity of the measurement with sample vibration was higher than that without sample vibration.
Both numerical analysis and experimental results confirm that 
this approach improves sensitivity, 
enabling characterization of high-resistance materials such as semiconductors.

\section*{Acknowledgments}
This work was supported by JSPS KAKENHI Grant Number 22H01498.

\section*{Reference}
\bibliographystyle{JJAP_by_Wakaya}

\end{document}